\documentclass[a4paper,11pt]{article}
\usepackage{pos}

\usepackage{amsmath,amsfonts,graphics,graphicx,amscd,amsfonts,epsfig,color}
\usepackage{subfig}
\usepackage{enumitem}
\usepackage{here}
\newcommand {\beq} {\begin{equation}}
\newcommand {\eeq} {\end{equation}}
\newcommand {\beqa}{\begin{eqnarray}}
\newcommand {\eeqa}{\end{eqnarray}}
\newcommand {\nn} {\nonumber}
\newcommand {\del} {\partial}

\newcommand {\Tr}{\mbox{Tr\,}}

\newcommand {\ee}{\mbox{e}}

\newcommand {\comm}[2]{\boxbc{#1, #2}}

\newcommand {\boxbc}[1]{\left[ #1 \right]}

\newcommand{\bbR}{{\mathbb R}}
\newcommand{\bbC}{{\mathbb C}}

\title{Signature change of the emergent space-time in the IKKT matrix model}
\ShortTitle{Signature change of the emergent space-time...}

\author*[a,b]{Jun Nishimura}

\affiliation[a]{Theory Center, Institute of Particle and Nuclear Studies,\\
High Energy Accelerator Research Organization (KEK),\\
1-1 Oho, Tsukuba, Ibaraki 305-0801, Japan}

\affiliation[b]{Department of Particle and Nuclear Physics, 
School of High Energy Accelerator Science,\\
Graduate University for Advanced Studies (SOKENDAI),\\
1-1 Oho, Tsukuba, Ibaraki 305-0801, Japan}

\emailAdd{jnishi@post.kek.jp}

\abstract{The IKKT matrix model (or the type IIB matrix model) is known as a promising candidate
  for a nonperturbative formulation of superstring theory in ten dimensions.
  As a most attractive feature, the model admits
  the emergence of (3+1)-dimensional space-time associated
  with the spontaneous breaking of the (9+1)-dimensional Lorentz symmetry.
  Numerical confirmation of such a phenomenon has been attempted for more
  than two decades. Recently it has been found that the sign problem,
  the main obstacle in simulating this model, can be overcome
  by the complex Langevin method. It has been shown that the
  Lorenzian version
  of the model is smoothly connected with the Euclidean version,
  in which the SO(10) symmetry is found to be spontaneously broken
  to SO(3).
  Here we propose to add a
  Lorentz invariant ``mass'' term to the original model
  and discuss a scenario that (3+1)-dimensional expanding space-time
  with Lorentzian signature
  appears at late times. Some numerical results supporting this scenario
  are presented.
}

\FullConference{%
  Corfu Summer Institute 2021 "School and Workshops on Elementary Particle Physics and Gravity"\\
  29 August - 9 October 2021\\
  Corfu, Greece
}

\begin{document}
\maketitle


\section{Introduction}


In view of the success of lattice gauge theory in understanding
the nonperturbative dynamics of QCD such as quark confinement and
the spontaneous chiral symmetry breaking,
it is quite natural to consider that
some kind of nonperturbative formulation of superstring theory
should play a crucial
role in understanding the expected dynamics
such as the compactification of six extra dimensions.
Indeed in 1996, the IKKT matrix model \cite{Ishibashi:1996xs}
(or the type IIB matrix model)
was proposed as
such a formulation.
Since then,
numerical studies have been performed by many people
in order to understand the dynamical properties of the model.

The model has ten bosonic $N\times N$ Hermitian matrices, whose
eigenvalue distribution
describes the ten-dimensional space-time in the large-$N$ limit.
It is therefore possible that
the eigenvalue distribution
collapses to a lower-dimensional manifold, which
implies that the ``compactification'' can occur dynamically
in this model.
When this happens,
the (9+1)D Lorentz symmetry of the model
has to be spontaneously broken.

There are quite a few pieces of evidence
that the IKKT model provides
a nonperturbative formulation of superstring theory.
In particular, direct connections to perturbative formulation
of superstring theory can be seen by considering the type IIB superstring 
theory in 10D.
First, the action of the model
can be regarded as a kind of matrix regularization of
the worldsheet action of type IIB superstring theory
in the Schild gauge \cite{Ishibashi:1996xs}.
Unlike the worldsheet formulation of superstring theory, however,
the matrix model is expected to be
a ``second quantization'' of superstrings
because multiple worldsheets appear naturally
in the matrix model as block-diagonal configurations,
where each block represents a single worldsheet
embedded into the 10-dimensional target space.
Second, under a few modest assumptions,
one can derive the string field Hamiltonian for type IIB superstring theory
from Schwinger-Dyson equations for the Wilson loop
operators, which play the role of creation and annihilation operators
of strings \cite{Fukuma:1997en}.
If this is true, the IKKT matrix model can reproduce
perturbative expansions in type IIB superstring theory to all orders.


In these connections to type IIB superstring theory,
one identifies the eigenvalues of the matrices $A_\mu$
as the target space coordinates.
This identification is suggested also by
the supersymmetry algebra of the model, in which the translation
that appears from the anti-commutator of supersymmetry generators
is identified with the shift symmetry
$A_\mu \mapsto A_\mu + \alpha_\mu {\bf 1}$
of the model, where $\alpha_\mu \in {\bf R}$.
It is also important to note that
the model has extended ${\cal N}=2$ supersymmetry
in ten dimensions,
which suggests that
the model should include gravity since it is known in field theory
that ${\cal N}=1$ supersymmetry is the maximal one that
can be achieved in ten dimensions without including gravity.


For many years, the IKKT matrix model was studied in
its SO(10) symmetric Euclidean version,
which is related to the Lorentzian version
by deforming the integration contour of the
10 bosonic matrices.
This contour deformation amounts
to multiplying some phase factors
to the temporal and spatial matrices,
which is allowed
since there is no singularity that one has to go through.
We can confirm explicitly that the Lorentzian version
is indeed equivalent to the Euclidean version
by measuring correlation functions, which are identical
up to some phase factors \cite{Hatakeyama:2021ake,Hatakeyama:2022ybs}.
Since
the emergent space-time in the Euclidean version has Euclidean signature,
the equivalence between the Lorentzian and Euclidean versions clearly
poses a big challenge in obtaining a Lorentzian space-time
in the IKKT model.
(See Refs.~\cite{Brahma:2021tkh,Steinacker:2021yxt,Klinkhamer:2021nyt}
for other recent developments on this model.)
%

Here we
overcome this situation
by adding a Lorentz invariant ``mass'' term to the original
model \cite{work_in_prog}.
The motivation for this term comes from our observation \cite{Hatakeyama:2019jyw}
that such a modified model have classical solutions
representing
space-time
with Lorentzian signature,
which exhibits expanding behavior
with any number of expanding directions.
If
the mass is large enough,
the path integral is expected to be
dominated by
one of such classical solutions.
The equivalence to the Euclidean model
is avoided
since the corresponding SO(10) invariant mass term will have
a phase factor $e^{\frac{3}{4} \pi i}$ with a negative real part,
which makes the integral divergent.
The mass
is sent to zero
after taking the large-$N$ limit
so that
the supersymmetry is restored.
In such a limit, we expect to obtain an inequivalent
model,
in which
(3+1)D expanding space-time with Lorentzian signature appears
at late times.\footnote{The Monte Carlo simulations
  in Refs.~\cite{Kim:2011cr,Ito:2013ywa,Ito:2015mxa,Ito:2017rcr,Azuma:2017dcb}
  avoided the sign problem by an approximation, which turned out to be
  unjustifiable \cite{Aoki:2019tby}. Moreover, the observed
  3D expanding space turned out to be represented by Pauli matrices \cite{Aoki:2019tby}.
  The present work is based on the CLM to overcome the sign problem \cite{Nishimura:2019qal},
  which revealed a new phase with the emergent space-time being continuous
  instead of having the Pauli-matrix structure \cite{Hirasawa:2021xeh}.}

Unfortunately, it is extremely hard to
perform Monte Carlo studies of these matrix models
due to the so-called sign problem
caused by the complex weight in the partition function.
In the Euclidean IKKT model, it comes from the Pfaffian
that is obtained by integrating out fermionic matrices,
while in the Lorentzian IKKT model, it comes from
the phase factor $\ee^{iS_{\rm b}}$ with the bosonic action $S_{\rm b}$.
If we treat the phase of the complex weight by reweighting,
huge cancellation among configurations with different phases occurs,
which makes the calculation impractical.
Recently the complex Langevin 
method (CLM) \cite{Parisi:1984cs,Klauder:1983sp}
has been attracting much attention
as a promising approach to this 
problem \cite{Aarts:2009uq,Aarts:2011ax,Nishimura:2015pba,Nagata:2015uga,Nagata:2016vkn}.
Here we use the same method in addressing the issues discussed
above using a technique \cite{Nishimura:2019qal}
which enables 
us to extract the time evolution \cite{Kim:2011cr} from matrices
generated by the CLM.


The rest of this article is organized as follows. 
In section \ref{sec:definition} we
discuss the problem in the original IKKT matrix model
that one cannot obtain space-time with Lorentzian signature.
In section \ref{sec:adding-mass}
we introduce the Lorentz invariant mass term
to solve the problem of the original model.
In section \ref{sec:extract-time-evolution} we discuss
how to extract the time evolution from matrix configurations.
In section \ref{sec:CLM} we discuss how we apply the CLM
to this model.
In section \ref{sec:expanding} we present our numerical results
for the bosonic model with the mass term,
which show the emergence of (1+1)D expanding space-time.
Section \ref{sec:summary} is devoted to a summary and discussions.
In particular, we speculate on the emergence of
(3+1)D expanding space-time
when the fermionic matrices are added.

\section{A problem in the original IKKT matrix model}
\label{sec:definition}

The action of the IKKT matrix 
model
is given by \cite{Ishibashi:1996xs}
\begin{eqnarray}
S & = & S_{{\rm b}}+S_{{\rm f}} \ ,
\label{eq:S_likkt}\\
S_{{\rm b}} & = & 
- \frac{1}{4g^{2}}{\rm Tr}
\left(\big[A_{\mu},A_{\nu}\big]
\big[A^{\mu},A^{\nu}\big]\right) \ ,
\label{eq:Sb}\\
S_{{\rm f}} & = & 
- \frac{1}{2g^{2}}{\rm Tr}
\left(\Psi_{\alpha}\left(\mathcal{C}
\Gamma^{\mu}\right)_{\alpha\beta}
\big[A_{\mu},\Psi_{\beta}\big]\right) \ ,
\label{eq:Sf-1}
\end{eqnarray}
where the bosonic variables
$A_{\mu}$ $\left(\mu=0,\ldots,9\right)$
and the fermionic variables
$\Psi_{\alpha}$ $\left(\alpha=1,\ldots,16\right)$
are $N\times N$ Hermitian matrices. 
$\Gamma^{\mu}$ are 10D gamma-matrices
after the Weyl projection and $\mathcal{C}$ is the charge conjugation
matrix. The ``coupling constant'' $g$ is merely a scale parameter
in this model since it can be absorbed by rescaling $A_{\mu}$ and
$\Psi_\alpha$ appropriately. 
In what follows, we set $g^2 = \frac{1}{N}$ without loss of generality.
The indices $\mu$ and $\nu$
are contracted using the Lorentzian metric 
$\eta_{\mu\nu}={\rm diag}\left(-1,1,\ldots,1\right)$.

The partition function 
of the Lorentzian IKKT matrix model can be written as \cite{Kim:2011cr}
\begin{equation}
Z=\int dA \, e^{iS_{{\rm b}}} \, {\rm Pf}\mathcal{M}\left(A\right)  \ ,
\label{Z-Likkt1}
\end{equation}
where ${\rm Pf}\mathcal{M}\left(A\right)$
is obtained by integrating out the fermionic matrices,
and it is a polynomial in $A$ that is known to take real values.
The ``$i$'' in front of the bosonic action
is motivated from the fact that
the string worldsheet metric should also have 
Lorentzian signature.
%
Note also that the bosonic action \eqref{eq:Sb}
can be written as 
\begin{eqnarray}
S_{\rm b}  =  
\frac{1}{4} N {\rm Tr}\left(F_{\mu\nu}F^{\mu\nu}\right)
 =  \frac{1}{4} N
\left\{ {\rm -2Tr}\left(F_{0i}\right)^{2}+
{\rm Tr}\left(F_{ij}\right)^{2}\right\} \ ,
\label{decomp-Sb}
\end{eqnarray}
where we have introduced
the Hermitian matrices $F_{\mu\nu}=i [A_{\mu},A_{\nu}]$.
Since the two terms
in the last expression
have opposite signs,
$S_{{\rm b}}$ is not positive semi-definite,
and it is not bounded from below.

Clearly the integral that appears in \eqref{Z-Likkt1} 
is not absolutely convergent.
In order to cure this problem,
we use Cauchy's theorem and deform the integration contour
for $A_\mu$ in
\eqref{Z-Likkt1}
as
\begin{align}
A_0 &= e^{- 3 s \pi i / 8} \tilde{A}_0 \ , \nn \\
A_i &= e^{s \pi i / 8} \tilde{A}_i  \ ,
\label{rotate-contour}
\end{align}
where $\tilde{A}_\mu$ are Hermitian.
This amounts to making the bosonic action
\begin{alignat}{3}
  S_{\rm b} &= N
\left\{ 
- \frac{1}{2} \ee^{- s \pi i / 2} \Tr (\tilde{F}_{0i})^2 
+ \frac{1}{4} \ee^{ s \pi i / 2} \Tr (\tilde{F}_{ij})^2 
\right\}
\label{sdef-action0}
\\
&= N \, \ee^{ s \pi i / 2}
\left\{ 
\frac{1}{2}  \Tr [ \ee^{- s \pi i/2 } \tilde{A}_0 , \tilde{A}_i ]^2
+ \frac{1}{4}  \Tr (\tilde{F}_{ij})^2 
\right\}  \ ,
\label{sdef-action}
\end{alignat}
where we have defined $\tilde{F}_{\mu\nu} =
i \, [\tilde{A}_\mu , \tilde{A}_\nu ]$.
The overall phase factor $\ee^{ s \pi i / 2}$ in \eqref{sdef-action}
can be identified
as the Wick rotation of the worldsheet coordinates,
whereas
the phase factor $\ee^{- s \pi i /2 }$ in front of
$\tilde{A}_0$
can be identified as the Wick rotation of the target space coordinates.
Similarly, the Pfaffian ${\rm Pf}\mathcal{M}\left(A\right)$ in
\eqref{Z-Likkt1}
is replaced by
${\rm Pf}\mathcal{M}(\ee^{- s \pi i /2} \tilde{A}_0, \tilde{A}_i)$ up to some
irrelevant constant phase factor.
In particular, at $s=1$, one obtains the Euclidean IKKT model,
for which the bosonic part $e^{iS_{\rm b}}$ in \eqref{Z-Likkt1}
becomes real positive for \eqref{sdef-action}
and the fermionic part
${\rm Pf}\mathcal{M}( -i \tilde{A}_0, \tilde{A}_i)$ becomes complex.
From \eqref{sdef-action0}, on the other hand, one finds
${\rm Im} S_{\rm b } > 0$ for generic
configurations\footnote{There is a subtlety due to the flat direction
    $S_{\rm b } =0 $
  corresponding to the configurations that
  satisfy $[\tilde{A}_\mu , \tilde{A}_\nu]=0$.
Despite this subtlety, the finiteness of the partition function
  \eqref{Z-Likkt1} is confirmed for $s=1$,
  which corresponds to the Euclidean IKKT
  model \cite{Krauth:1998xh,Austing:2001pk}.
  We consider that
  the proof in Ref.~\cite{Austing:2001pk} can be extended to $0<s<2$.}
if $0<s<2$, which suggests
that the model \eqref{Z-Likkt1} becomes well defined in that
region.
Therefore, one can define the Lorentzian model by taking the
$s \rightarrow 0$ limit\footnote{The author
  would like to thank Yuhma Asano for pointing this out to him in 2018.}.

The model defined in this way is actually equivalent
to the Euclidean IKKT model.
For instance, we obtain \cite{Hatakeyama:2021ake,Hatakeyama:2022ybs}
\begin{align}
\left\langle \frac{1}{N} \Tr (A_0)^2 
\right\rangle_{\rm L}
&= e^{-\frac{3\pi}{4}i}
\left\langle \frac{1}{N} \Tr (\tilde{A}_0)^2 
\right\rangle_{\rm E} \ ,
\label{equiv-A0}
\\
\left\langle \frac{1}{N} \Tr (A_i)^2 
\right\rangle_{\rm L}
&= e^{\frac{\pi}{4}i}
\left\langle \frac{1}{N} \Tr (\tilde{A}_i)^2 
\right \rangle_{\rm E} \ ,
\label{equiv-Ai}
\end{align}
where the suffixes ``L'' and ``E'' imply
that the expectation values are defined in the
Lorentzian model and the Euclidean model, respectively.
One can prove that the expectation values 
$\left\langle \frac{1}{N} \Tr  (\tilde{A}_0)^2 
\right\rangle_{\rm E}$
and 
$\left\langle \frac{1}{N} \Tr  (\tilde{A}_i)^2 
\right \rangle_{\rm E}$
are real positive using the fact that
the Pfaffian becomes complex conjugate
under the parity transformation
$\tilde{A}_\mu \mapsto  - \tilde{A}_\mu$
for some $\mu$.
This means that 
$\left\langle \frac{1}{N} \Tr (A_0)^2 
\right\rangle_{\rm L}$
and 
$\left\langle \frac{1}{N} \Tr (A_i)^2 
\right\rangle_{\rm L}$
have the phase factors $e^{-\frac{3\pi}{4}i}$ and
$e^{\frac{\pi}{4}i}$, respectively.
Thus the space-time that appears dynamically in the Lorentzian
model has Euclidean signature.

The Euclidean IKKT model has been studied recently
by the complex Langevin method \cite{Anagnostopoulos:2020xai},
and the rotational SO(10) symmetry is found to be spontaneously broken
down to SO(3) as suggested earlier by the Gaussian expansion
method \cite{Nishimura:2011xy}.
While this is certainly an interesting dynamical property
of the Euclidean IKKT model, its relevance to our real world is unclear.


\section{Adding a Lorentz invariant ``mass'' term}
\label{sec:adding-mass}

In order to overcome this situation,
we propose to add a Lorentz invariant ``mass'' term
to the original model \cite{work_in_prog}.
Namely, we add to the bosonic action \eqref{decomp-Sb},
a quadratic term
\begin{eqnarray}
S_{\rm m}  =  
-\frac{1}{2} N  \gamma {\rm Tr}\left(A_{\mu} A^{\mu}\right)
= \frac{1}{2} N  \gamma \left\{  {\rm Tr} (A_0)^2
- {\rm Tr} (A_i)^2  \right\} \ ,
\label{mass-term}
\end{eqnarray}
where $\gamma > 0$ is a parameter which is sent to zero after
taking the large-$N$ limit.

The motivation for this mass-deformed model defined above
comes from our observation \cite{Hatakeyama:2019jyw}
that the classical equation of motion\footnote{The
  mass term was introduced
  originally 
  to represent the
  effects of the infrared cutoff used in the simulation \cite{Kim:2011cr}.}
\begin{equation}
\label{eq: EOM}
\comm{A^\nu}{\comm{A_\nu}{A_\mu}}-\gamma A_\mu = 0
\end{equation}
derived from it
has infinitely many solutions with Hermitian $A_\mu$,
which represent space-time with Lorentzian signature.
Moreover, these solutions
exhibit expanding behavior generically
with any number of expanding directions.
If the coefficient $\gamma$ of the mass term is large enough,
the path integral is expected to be
dominated by
one of such classical solutions.\footnote{This can be understood
  by rescaling $A_\mu = \sqrt{\gamma} \tilde{A}_\mu$, which
  makes the total action proportional to $\gamma^2$.
  Hence at large $\gamma$, the path integral is dominated
  by some saddle-point configuration satisfying
  $\frac{\del S}{\del A_\mu}=0$, which is nothing but \eqref{eq: EOM}.}
Our numerical results presented below
suggest that the expanding behavior continues
longer as we decrease $\gamma$ although there is a transition
to the ``Euclidean phase'' at some critical $\gamma_{\rm c}(N)$ for finite $N$.
It is conceivable that $\gamma_{\rm c}(N)\rightarrow 0 $ in the large-$N$
limit since the expanding phase is entropically
favored in that limit
compared to the Euclidean phase, in which the extent of space-time
remains finite.
Thus we expect to obtain a space-time with Lorentzian signature
and expanding behavior if we take
the $\gamma \rightarrow 0$ limit after the large-$N$ limit.

The reason why we can avoid
the problem of the original model discussed in the previous section
can be understood as follows.
Upon rotation of the integration contour \eqref{rotate-contour},
one obtains
\begin{eqnarray}
  S_{\rm m}
&=& \frac{1}{2} N  \gamma 
\left\{  \ee^{- 3 s \pi i / 4} {\rm Tr} (\tilde{A}_0)^2
- \ee^{s \pi i / 4} {\rm Tr} (\tilde{A}_i)^2  \right\} \ .
\label{mass-term-rotated}
\end{eqnarray}
Unlike \eqref{sdef-action0}, one finds that
${\rm Im} S_{\rm m} <  0$ for $0<s<\frac{4}{3}$,
which implies that
the bosonic part $e^{iS_{\rm m}}$ in
the partition function
can become arbitrarily large in magnitude for \eqref{mass-term-rotated}.
Therefore, one cannot 
make the model well defined by simply introducing $s>0$
and taking the $s \rightarrow 0$ limit.
In particular, at $s=1$, one obtains
\begin{eqnarray}
  S_{\rm m}
&=& \frac{1}{2} N \gamma \, i \, \ee^{ 3 \pi i / 4}
\left\{   {\rm Tr} (\tilde{A}_0)^2 + {\rm Tr} (\tilde{A}_i)^2  \right\} \ ,
\label{mass-term-rotated-Euclid}
\end{eqnarray}
which
makes the corresponding Euclidean model ill defined
due to the phase factor $\ee^{ 3 \pi i / 4}$ with a negative real part
in front of the SO(10) invariant mass term.


In order to make the mass-deformed IKKT model well defined
for finite $\gamma$ and finite $N$,
we can think of introducing
some convergence factor in \eqref{mass-term} as
\begin{eqnarray}
S_{\rm m}^{(\varepsilon)}  
= \frac{1}{2} N  \gamma \left\{  e^{i\varepsilon} {\rm Tr} (A_0)^2
- e^{- i\varepsilon} {\rm Tr} (A_i)^2  \right\} \ .
\label{mass-term-epsilon}
\end{eqnarray}
We deform the integration contour as \eqref{rotate-contour}
and take the $s\rightarrow 0$ limit before
we take the $\varepsilon \rightarrow 0$ limit.




Let us emphasize that the sign of the mass term \eqref{mass-term}
is crucial.
For $\gamma<0$, we have
${\rm Im} S_{\rm m} > 0$ for $0<s<\frac{4}{3}$,
which enables us to connect the theory to the Euclidean
model with the mass term \eqref{mass-term-rotated-Euclid},
which is well defined.
Therefore the situation is qualitatively the same as in the $\gamma=0$ case,
and nothing dramatic happens.

It is also known \cite{Hatakeyama:2019jyw} that
the classical equation of motion \eqref{eq: EOM}
does not have expanding solutions for $\gamma<0$.
When $\gamma=0$, 
the classical equation of motion \eqref{eq: EOM}
is satisfied if and only if all the matrices are commutative;
i.e.,\ $[A_\mu , A_\nu]=0$ as is proved in Appendix A of
ref.~\cite{Steinacker:2017vqw}.

The mass term can be interpreted as the cosmological constant
in the Einstein equation, which is derived
from the IKKT matrix model \cite{Hanada:2005vr}.
Incidentally, the mass term is introduced
in obtaining interesting classical solutions
in Refs.~\cite{Kim:2011ts,Kim:2012mw,Steinacker:2017vqw,Steinacker:2017bhb,Sperling:2019xar,Steinacker:2021yxt}.
See also
Refs.\cite{Chaney:2015ktw,Chaney:2015mfa,Chaney:2016npa,Stern:2018wud}
for related work, which discuss the signature change in the IKKT type of matrix models
from a different viewpoint.

\section{How to extract the time evolution}
\label{sec:extract-time-evolution}

Let us explain how we can extract the time evolution from
matrix configurations following ref.~\cite{Kim:2011cr}.
For that, we use
the ${\rm SU}\left(N\right)$ symmetry of the model
to bring the temporal matrix $A_{0}$ into the diagonal form
\begin{equation}
A_{0}={\rm diag}\left(\alpha_{1},\ldots,\alpha_{N}\right)\ ,
\quad \quad
{\rm where~} \alpha_{1}< \ldots < \alpha_{N} \ .
\label{eq:diagonal gauge}
\end{equation}
By ``fixing the gauge'' in this way,
we can rewrite the partition function (\ref{Z-Likkt1}) as
\beqa
\label{gauge-fixing}
Z &=& \int  \prod_{a=1}^{N}d\alpha_{a}\,
\Delta (\alpha)^2  \int dA_i \, e^{i(S_{\rm b}+S_{\rm m})} \, 
       {\rm Pf}\mathcal{M}\left(A\right) \ ,
       \label{A0diag}
\\
\Delta (\alpha) &\equiv &
\prod_{a>b}^{N}
\left(\alpha_{a}-\alpha_{b}\right) \ ,
\label{VDM}
\eeqa
where $\Delta(\alpha)$ is the van der Monde determinant.
The factor $\Delta (\alpha)^2$ 
in (\ref{gauge-fixing})
appears from the Fadeev-Popov procedure
for the gauge fixing, and it acts as a repulsive potential 
between $\alpha_a$.

We can extract a time-evolution
from matrix configurations of $A_\mu$.
A crucial observation is that 
the spatial matrices $A_{i}$ have
a band-diagonal structure
in the SU($N$) basis in which $A_{0}$ 
has the diagonal form (\ref{eq:diagonal gauge}).
See Fig.~\ref{fig:gamma=3} (Right).
More precisely, there exists some integer $n$ such that
the elements of spatial matrices
$\left(A_{i}\right)_{ab}$ for $\left|a-b\right|>n$ are 
much smaller than those for $\left|a-b\right|\leq n$.
Based on this observation,
we may naturally consider $n\times n$ submatrices of $A_i$,
\begin{equation}
\left(\bar{A}_{i}\right)_{IJ}\left( t_\nu \right)
\equiv\left(A_{i}\right)_{\nu+I,\nu+J} \ ,
\label{eq:def_abar}
\end{equation}
where $I,J=1,\ldots , n$, $\nu=0,1,\ldots , N-n$,
and $t_\nu$ is defined by 
\begin{align}
  t_\nu &= \sum_{\rho=1}^{\nu}
  | \bar{\alpha}_{\rho}  - \bar{\alpha}_{\rho-1}  | \ , \\
  \bar{\alpha}_\nu & = \frac{1}{n}\sum_{I=1}^{n} \langle \alpha_{\nu+I} \rangle \ .
\label{eq:def_t}
\end{align}
We interpret the $\bar{A}_i(t)$ 
as representing the state of the universe at time $t$.
Note that $\bar{\alpha}_\nu \in \bbC$ in general,
since the weight in \eqref{A0diag} is complex.
The appearance of real time implies that
$\bar{\alpha}_{\nu}  - \bar{\alpha}_{\nu-1} \in \bbR$.


Using $\bar{A}_{i}(t)$,
we can define, for example,
the extent of space at time $t$ as 
\begin{equation}
R^{2}\left(t\right)=
\left\langle \frac{1}{n}{\rm tr}\sum_{i}
\left(\bar{A}_{i}\left(t\right)\right)^{2}\right\rangle \ ,
\label{eq:def_rsq}
\end{equation}
where
${\rm tr}$ represents
a trace over the $n\times n$ submatrix.
Since $R^{2}\left(t\right) \in \bbC$ in general,
let us define
\begin{align}
  R^2(t) = e^{2i \theta_{\rm s} (t)} |R^2(t)| \ .
  \label{R2-theta_s-def}
\end{align} 
The appearance of real space implies that $\theta_{\rm s}(t) = 0 $.

We also define
the ``moment of inertia tensor'' 
\begin{equation}
T_{ij}\left(t\right)
=\frac{1}{n}{\rm tr}
\Big(\bar{A}_{i}(t) \bar{A}_{j}(t)\Big) \ ,
\label{eq:def_tij}
\end{equation}
which is a $9\times9$ real symmetric matrix
since $\bar{A}_i(t)$ is Hermitian. 
The eigenvalues of $T_{ij}\left(t\right)$,
which we denote by $\lambda_{i}\left(t\right)$ with the order
\begin{equation}
\lambda_{1}\left(t\right)>\lambda_{2}
\left(t\right)> \ldots > \lambda_{9}\left(t\right) \ ,
\label{lambda-def}
\end{equation}
represent the spatial extent in each of 
the nine directions at time $t$.
Note that the expectation values 
$\left\langle \lambda_{i}\left(t\right)\right\rangle \in \bbC$
tend to be equal in the large-$N$ limit if the SO(9) symmetry is
not spontaneously broken.
If some of the eigenvalues
$\left\langle \lambda_{i}\left(t\right)\right\rangle$ 
($i=1, \ldots , d$)
have significantly larger modulus than the rest,
%
it implies that $d$-dimensional space appears dynamically.

\section{Applying the complex Langevin method}
\label{sec:CLM}

In this section, we review the CLM and discuss how we 
apply it to the model \eqref{A0diag}.
%
\subsection{Brief review of the CLM}
\label{sec:CLM-review}
%
%
Let us consider a system
\begin{align}
  Z=\int dx \, e^{-S(x)}
\label{original-theory}
\end{align}
of $N$ real variables $x_k$ ($k=1 ,\ldots , N$)
as a simple example. 
Here the action $S(x)$
is a complex-valued function, 
which causes the sign problem.



In the CLM, 
the original real variables $x_k$ 
are complexified as $x_k\to z_k=x_k+iy_k\in \mathbb{C}$
and one considers a fictitious time evolution of
the complexified variables $z_k$ using
the complex Langevin equation
given, in its discretized form, by
\begin{align}
z^{(\eta)}_k(t+\epsilon)=z^{(\eta)}_k(t)
+\epsilon \, v_k(z^{(\eta)}(t))+\sqrt{\epsilon} \, \eta_k(t) \ ,
\label{Langevin}
\end{align}
where $t$ is the fictitious time with a stepsize $\epsilon$.
The second term $v_k(z)$ on the right-hand side is called 
the drift term, which is defined by
holomorphic extension of the one
\begin{align}
  v_k(x)=
   -  \frac{\partial S(x)}{\partial x_k}
\end{align}
for the real variables $x_k$.
The variables $\eta_k(t)$ appearing on the right-hand side of 
eq.~(\ref{Langevin})
are a real Gaussian noise 
with the probability distribution
$\propto e^{-\frac{1}{4}\sum_t\eta_k(t)^2}$,
which makes the time-evolved variables $z^{(\eta)}_k(t)$ stochastic.
The expectation values
with respect to the noise $\eta_k(t)$ 
are denoted as $\langle \cdots \rangle_\eta$ in what follows.

Let us consider the expectation value of 
an observable $\mathcal O(x)$.
In the CLM, one computes the expectation value of the 
holomorphically extended observable
$\mathcal O(z)$ for complexified variables $z_k$.
Then, the correct convergence of the CLM implies the equality
\begin{align}
  \lim_{t\to \infty}\lim_{\epsilon \to 0}
\Big\langle \mathcal O(z^{(\eta)}(t)) \Big\rangle_\eta
=\frac{1}{Z}\int dx \, \mathcal O (x) \, 
e^{-S(x)}
\ ,
  \label{key}
\end{align}
where the right-hand side is the expectation value 
of $\mathcal O(x)$ in the original theory (\ref{original-theory}).
A proof of eq.~\eqref{key} was given
in refs.~\cite{Aarts:2009uq,Aarts:2011ax}, where the notion of
the time-evolved observable $\mathcal O (z;t)$ plays a crucial role.
In particular, it was pointed out that the integration by parts 
used in the argument cannot be justified 
when the probability distribution of
$z_k$
that appears during the simulation
falls off slowly
in the imaginary direction.

While this argument provided theoretical understanding of the cases
in which the CLM gives wrong results, the precise condition on the 
probability distribution was not specified.
Furthermore, there is actually a subtlety
in defining the time-evolved observable $\mathcal O (z;t)$.
Recently ref.~\cite{Nagata:2016vkn} provided
a refined argument for justification
of the CLM, which showed that
the probability for the drift term $v_k(z)$ to become large
has to be suppressed strongly enough.
More precisely the histogram of the magnitude of the 
drift term should fall off exponentially or faster.
This criterion tells us whether the results obtained by the CLM
are reliable or not.


\subsection{Applying the CLM to
 the bosonic IKKT model with the mass term}
\label{sec:applying-CLM}

Let us 
apply the CLM to the model \eqref{A0diag}
following ref.~\cite{Nishimura:2019qal}.
From now on, we omit the Pfaffian
and consider the bosonic model for simplicity.

The first step of the CLM is to complexify the real variables.
As for the spatial matrices $A_i$, 
we simply treat them
as general complex matrices instead of Hermitian matrices.
As for the temporal matrix $A_0$, which is diagonalized as
(\ref{eq:diagonal gauge}),
we have to take into account the ordering of the eigenvalues.
For that purpose,
we make the change of variables as
\begin{alignat}{3}
\alpha_1 = 0 \ , \quad
\alpha_2 = \ee^{\tau_1} \ , \quad
\alpha_3 = \ee^{\tau_1} + \ee^{\tau_2} \ ,
\quad
\ldots  \ , \quad 
\alpha_N = \sum_{a=1}^{N-1} 
\ee^{\tau_a}  
  \label{eq:alpha-tau}
\end{alignat}
so that the ordering is implemented automatically,
and then complexify $\tau_a$ ($a=1 ,\ldots , N-1$).
Using the shift symmetry
$A_0 \mapsto A_0 + {\rm const}. {\bf 1}$ of the original IKKT action,
we make a shift $A_0  \rightarrow  A_0 - \frac{1}{N} \Tr A_0$
so that $A_0$ becomes traceless in what follows.

The effective action that appears in the Boltzmann weight
$e^{- S_{\rm eff}}$ reads
\begin{alignat}{3}
S_{\rm eff}  
=&  - \frac{1}{4} i N \Big\{ 
2 \, \Tr [A_0 , A_i]^2 
 - \Tr [A_i , A_j]^2
\Big\}
-   \frac{1}{2} i N  \gamma \Big\{  {\rm Tr}
({A}_0)^2 - {\rm Tr} (A_i)^2  \Big\} 
\nn \\
&  - \log \Delta(\alpha)  
- \sum_{a=1}^{N-1}\tau_a \ ,
\label{eq:eff_action}
\end{alignat}
where the last term comes from the Jacobian associated
with the change of variables (\ref{eq:alpha-tau}).
The complex Langevin equation is given by
\begin{alignat}{3}
\frac{d\tau_a}{dt}&=
 - \frac{\del S_{\rm eff}}{\del \tau_a} + \eta_a(t)  \ ,   \nn \\
\frac{d (A_i)_{ab}}{dt}&=
 - \frac{\del S_{\rm eff}}{\del (A_i)_{ba}} 
+ (\eta_i)_{ab}(t)  \ ,
  \label{eq:cle}
\end{alignat}
where the $\eta_a(t)$ in the first equation 
are random real numbers
obeying the probability distribution
$\exp (- \frac{1}{4}\int dt \sum_a \{ \eta_a(t) \}^2 )$
and the $\eta_i(t)$ in the second equation 
are random Hermitian matrices
obeying the probability distributions 
$\exp (- \frac{1}{4}\int dt \sum_{i} \Tr  \{ \eta_i(t) \}^2 )$.

The expectation values of observables can be calculated by 
defining them holomorphically for complexified $\tau_a$ and $A_i$
and taking an average using the configurations generated by
solving the discretized version of (\ref{eq:cle}) 
for sufficiently long time.
In order for this method to work, the probability distribution
of the drift terms, namely the first terms
on the right-hand side of (\ref{eq:cle}), has to fall off 
exponentially \cite{Nagata:2016vkn}.
We have checked that this criterion is indeed satisfied
for all the values of parameters used in this paper.

\begin{figure}[t]
\centering
\includegraphics[width=7cm]{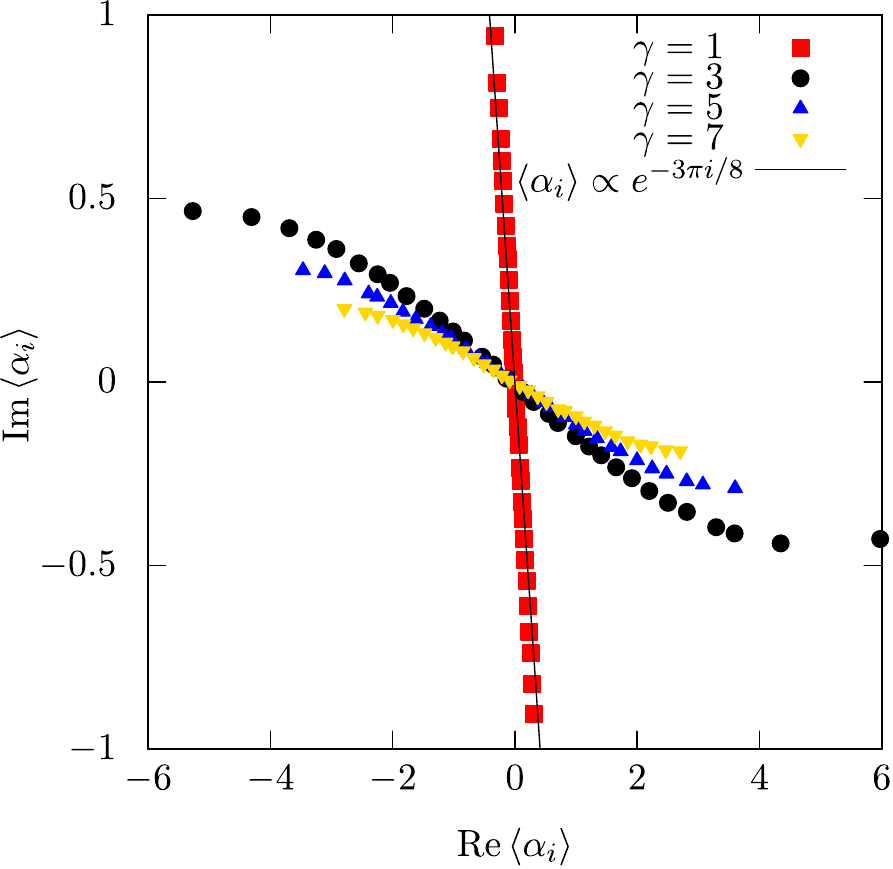}
\caption{The expectation values $\langle \alpha_i \rangle$
  are plotted in the complex plane for $\gamma=1,3,5,7$.
The solid line represents the prediction \eqref{equiv-A0}
for $\gamma=0$ obtained from the equivalence to the Euclidean IKKT model.
}
\label{fig:alpha}
\end{figure}

\section{Results for the bosonic IKKT model with the mass term}
\label{sec:expanding}

In this section we present our preliminary results for
the bosonic IKKT model \eqref{A0diag} with the mass term \cite{work_in_prog}.
We choose the matrix size to be $N=32$
and set $\varepsilon = 0$ in \eqref{mass-term-epsilon}
for simplicity.

When the mass term is absent $\gamma=0$,
we obtain the results equivalent to the Euclidean model
as in \eqref{equiv-A0} and \eqref{equiv-Ai}.
The complex Langevin simulation is completely stable.
This is understandable
since for $\gamma=0$ the simulation can find the
contour deformation by itself ending up in simulating
the Euclidean model, which is free from
the sign problem.\footnote{Note that this is no more the case if we incorporate
  the Pfaffian in the simulation.}

When we start our simulation with large $\gamma$, however,
the simulation turns out to be unstable,
which forces us to use some trick to obtain meaningful results.
Let us note here that
one of the classical solutions
represented by Hermitian $A_\mu$ is expected to
dominate at large $\gamma$.
Therefore, we insert a procedure
\begin{align}
  A_i \mapsto \frac{1}{1+\eta} \left( A_i  + \eta A_i^\dagger \right)
  \quad \quad \mbox{for~$i=1, \ldots , 9$}
  \label{partial-Hermitizing}
\end{align}
after each Langevin step to stabilize the
simulation,
which is similar in spirit to
  the dynamical stabilization proposed in the complex Langevin
  simulation of finite density QCD \cite{Attanasio:2018rtq}.
For $\eta=1$, this amounts to Hermitizing $A_i$
after each Langevin step, whereas $\eta=0$ corresponds
to doing nothing.
We tried to decrease $\eta$ as much as possible,
and found that the simulation is stable
for $\eta \gtrsim 0.001$ and the results do not depend much
on $\eta$ within the region $0.001 \lesssim \eta \lesssim 0.01$.
In what follows, we present our results for $\eta=0.01$.

After we obtain a thermalized configuration for $\gamma=7$ in this way,
we decrease $\gamma$ adiabatically and obtain results for smaller $\gamma$.
At $\gamma \sim 2.5$, there is a drastic change in the results.
The results for $\gamma=1$ are close to those for $\gamma=0$,
and
we set $\eta=0$ since the technique \eqref{partial-Hermitizing}
is not only needless but also unjustifiable since the configurations
are not close to Hermitian in this Euclidean phase.

In Fig.~\ref{fig:alpha}, we plot our results
for $\langle \alpha_i \rangle$
in the complex plane
for $\gamma=1,3,5,7$.
Note that the aspect ratio is chosen as $1:6$.
For $\gamma=7$, the distribution
of $\langle \alpha_i \rangle$ is close to the real axis,
which is consistent with the fact that
one of the classical solutions
represented by Hermitian $A_\mu$ dominates
at large $\gamma$.
As $\gamma$ becomes smaller,
we observe that
the distribution moves away from the real axis,
but the flat region at both ends extends,
suggesting the emergence of real time in that region.
The results for $\gamma=1$ are close to the prediction
\eqref{equiv-A0}
for $\gamma=0$ obtained from the equivalence to the Euclidean model,
and they are qualitatively different from the results for $\gamma \ge 3$,
suggesting a first order phase transition\footnote{This is also suggested
  by the existence of hysteresis; starting from a thermalized configuration
  at $\gamma=0$ and increasing $\gamma$ adiabatically, we find
  that the system is in the Euclidean phase even for $\gamma \gtrsim 2.5$.}
at some $\gamma = \gamma_{\rm c}(N)$,
which lies around $2.5$ for the present matrix size $N=32$.
From our preliminary results
for $N=64$, the (lower)
critical point $\gamma_{\rm c}(N)$ seems to decrease for larger $N$.

\begin{figure}[t]
\centering
\includegraphics[width=7cm]{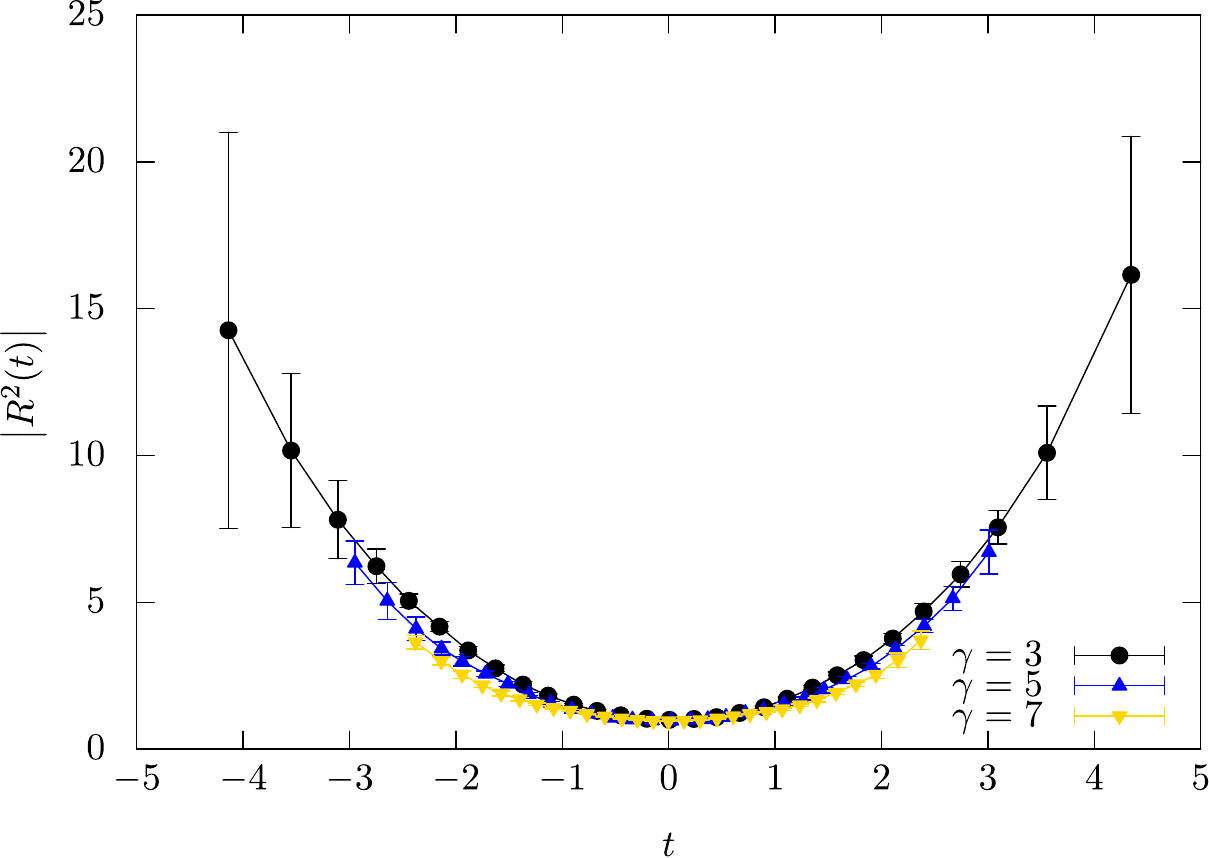}
\includegraphics[width=7cm]{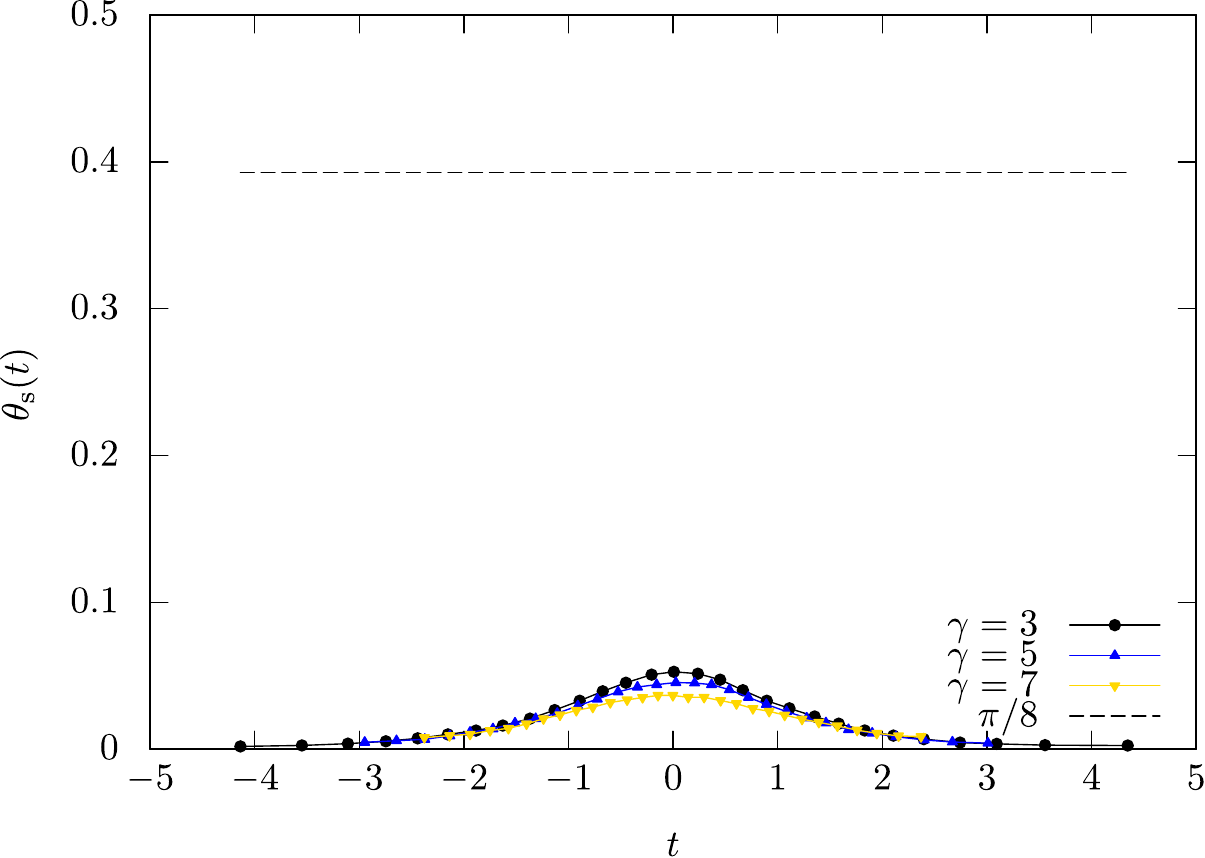}
\caption{(Left) The extent of space $|R^2(t)|$ is plotted
  against time $t$ for $\gamma=3,5,7$. (Right) The complex phase $\theta_{\rm s}(t)$
  of the space is plotted against time $t$ for $\gamma=3,5,7$.
  The dashed line $\theta_{\rm s}(t)=\frac{\pi}{8}$
  represents the prediction \eqref{equiv-Ai}
for $\gamma=0$ obtained from the equivalence to the Euclidean model.
}
\label{fig:Rsq-theta_s}
\end{figure}

In Fig.~\ref{fig:Rsq-theta_s},
we plot $|R^2(t)|$ (Left) and $\theta_{\rm s}(t)$ (Right)
defined by \eqref{R2-theta_s-def}
for $\gamma \ge 3$.
The block size used in defining $\bar{\alpha}_\nu$ in \eqref{eq:def_t}
and $\bar{A}_i(t)$ in \eqref{eq:def_abar} is chosen to be $n=4$.
From the left panel, we find that the expanding behavior
of $|R^2(t)|$ is analogous to that
observed for classical solutions \cite{Hatakeyama:2019jyw}.
Scaling behavior is observed for different values of $\gamma$,
and decreasing $\gamma$ results in extending the time direction
and hence the space becomes larger at the end time.
From the right panel, we find that the space becomes real
at late times, while the phase $\theta_{\rm s}(t)$ becomes positive
near $t \sim 0$.
Emergence of the real space-time at late times observed even for $\gamma=3$
can be understood as a consequence of classicalization
since the value of the action increases with the expansion \cite{Kim:2012mw}.

Figures \ref{fig:alpha} and \ref{fig:Rsq-theta_s}
exhibit symmetries around $t=0$, which is due to the symmetry
of the model \eqref{A0diag} under $A_0 \mapsto - A_0$.
The behavior is reminiscent of bouncing cosmology.

\begin{figure}[t]
\centering
\includegraphics[width=7cm]{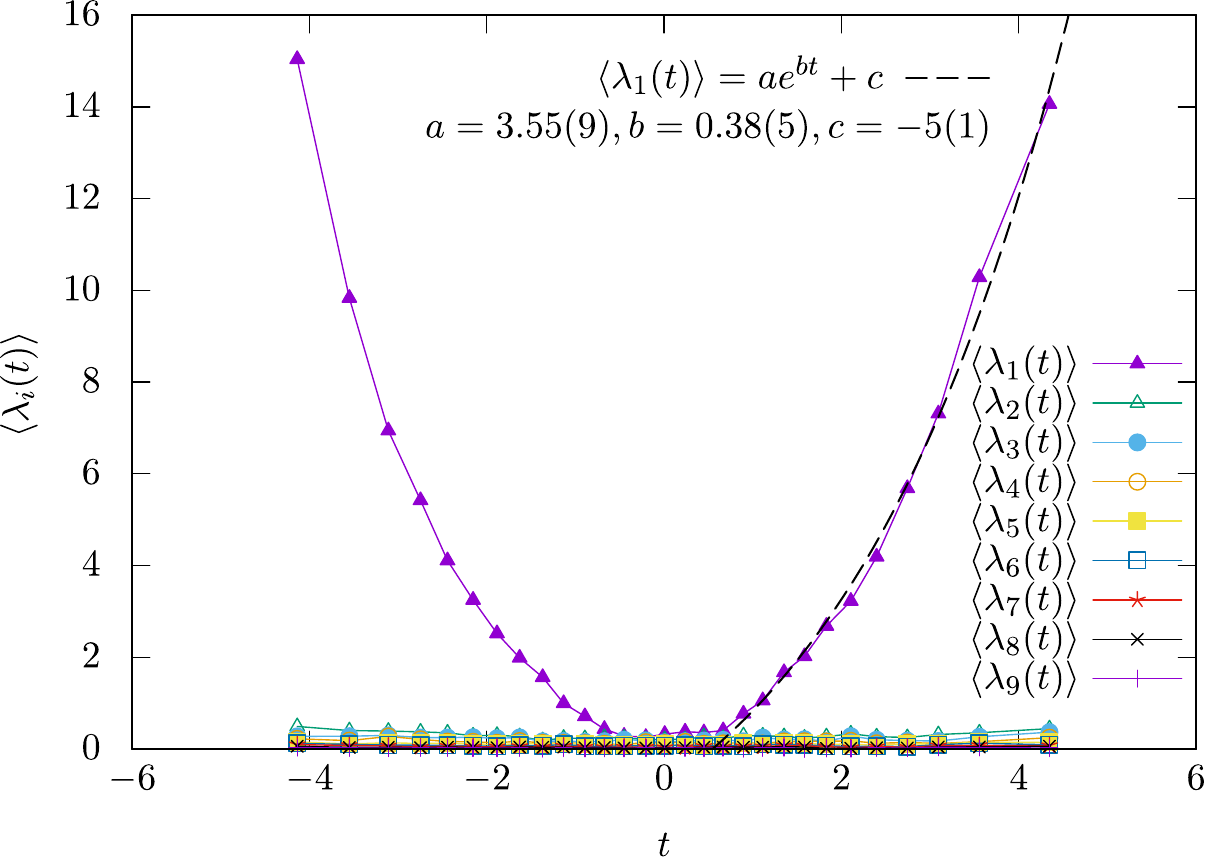}
\includegraphics[width=7cm]{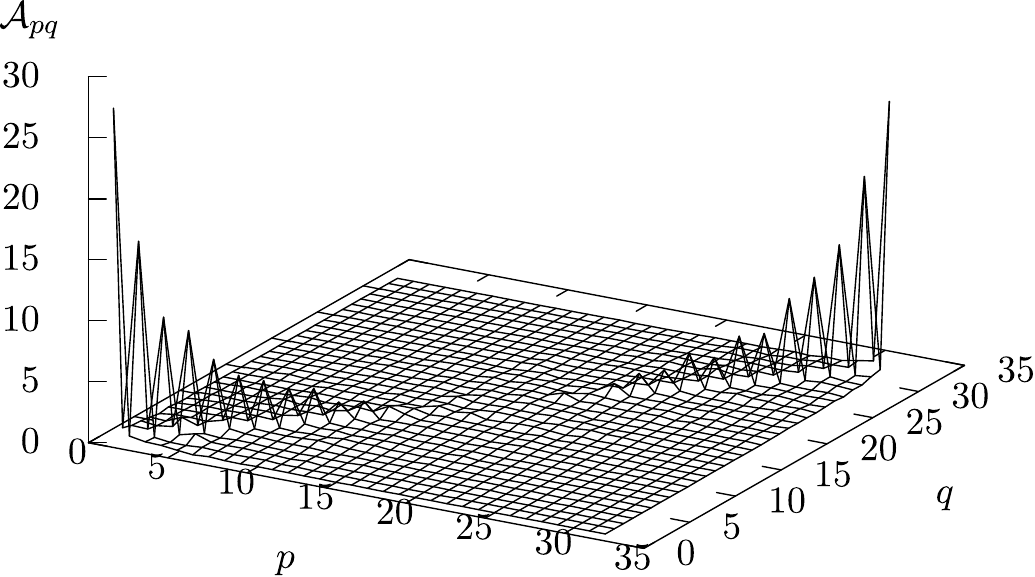}
\caption{(Left) The expectation values $\langle \lambda_i(t) \rangle$
  are plotted against $t$ for $\gamma=3$.
  The dashed line represents a fit of
  $\langle \lambda_1 (t) \rangle$ 
  to the behavior $\langle \lambda_1 (t) \rangle = a e^{b t} + c $,
where $a=3.55(9)$, $b=0.38(5)$ and $c=-5(1)$.
(Right) The magnitude ${\cal A}_{pq}$
of each element of the spatial matrices
is plotted against $p$ and $q$ for $\gamma=3$.}
\label{fig:gamma=3}
\end{figure}

Let us also look at the order parameter for the SSB of
SO(9) symmetry for $\gamma \ge 3$.
It is not straightforward to calculate the expectation values
of \eqref{lambda-def} by the CLM since they cannot be regarded as
holomorphic functions of $\tau_a$ and $A_i$.
Here we estimate them 
by defining the ``moment of inertia tensor'' \eqref{eq:def_tij}
using only the Hermitian part of $\bar{A}_{i}(t)$,
which is expected to be a good approximation
according to our results in Fig.~\ref{fig:Rsq-theta_s} (Right).
The results for $\gamma=3$ are shown in Fig.~\ref{fig:gamma=3} (Left).
We observe that only one direction expands and
the other directions remain small.
Thus the expanding space is actually one dimensional.
We can fit our data for the largest eigenvalue
$\langle \lambda_1 (t) \rangle$ to an exponential behavior,
which shows that our data are consistent with an exponential expansion.

Finally, let us confirm that
the spatial matrices $A_i(t)$ 
has a band diagonal structure, which is important
in defining the submatrices \eqref{eq:def_abar}.
For that, we plot
\begin{align}
  {\cal A}_{pq} = \frac{1}{9} \sum_{i=1}^9 |(A_i)_{pq}|^2
\end{align}
against $p$ and $q$ for $\gamma=3$
in Fig.~\ref{fig:gamma=3} (Right).
We find that the off-diagonal elements are quite small.
Similar behaviors are observed for $\gamma=5,7$,
which justifies our choice $n=4$ of the block size for $\gamma \ge 3$.
Such band-diagonal structure is not observed for $\gamma \le 2$.

\section{Summary and discussions}
\label{sec:summary}


In this article, we have discussed the signature
of the emergent space-time
in the IKKT matrix model,
which was proposed as a nonperturbative formulation of superstring theory.
A naive definition of the model leads to a space-time with Euclidean
signature. In order to avoid this, we have proposed to add
a Lorentz invariant mass term to the original action.
We investigate the bosonic IKKT model with the mass term by the CLM.
When the mass parameter $\gamma$ is large enough, the path integral is dominated
by one of the classical solutions
with Lorentzian signature and expanding behavior.
As $\gamma$ is decreased, the extent of the emergent time
increases and the emergent space at the end time becomes larger.
The signature of the space-time is Lorentzian at late times,
while it seems to deviate from Lorentzian towards Euclidean at early times.
The expansion at late times is consistent with an exponential behavior,
and we also observed that only one out of nine spatial directions
expands.
We speculate that an expanding space-time
with Lorentzian signature emerges at late times
even
in the $\gamma \rightarrow 0$ limit
after taking the large-$N$ limit.

The mechanism for the appearance of the (1+1)D space-time
may be understood from the bosonic action \eqref{decomp-Sb}.
Since the spatial direction expands exponentially,
  the $\Tr [A_i , A_j]^2$ term becomes dominant.
  The fluctuation of this term can be made small
  by having only one expanding direction.

As a future prospect, it would be important to include the effect
of fermionic matrices, which is represented by the Pfaffian in \eqref{Z-Likkt1}.
It is known that the Pfaffian vanishes if we set $A_\mu$
to zero except for two of them \cite{Krauth:1998xh,Nishimura:2000ds}.
Therefore the (1+1)D space-time observed in our simulation
of the bosonic model is strongly suppressed by the Pfaffian.
Considering that the expansion of space is exponential with respect to
time, it is conceivable that the Pfaffian favors the emergence
of three exponentially extended spatial directions.
Such effects are already confirmed in the Euclidean IKKT model,
in which one indeed obtains three extended
directions \cite{Nishimura:2011xy,Anagnostopoulos:2020xai}.

\section*{Acknowledgements}

The author would like to thank 
K.N.~Anagnostopoulos, Y.~Asano, T.~Azuma, K.~Hatakeyama, M.~Hirasawa, Y.~Ito,
F.~Klinkhamer, S.K.~Papadoudis, H.~Steinacker and A.~Tsuchiya for
valuable discussions.
This research was supported by MEXT as
``Program for Promoting Researches on the Supercomputer Fugaku'' (Simulation
for basic science: from fundamental laws of particles to creation of nuclei,
JPMXP1020200105) and JICFuS. This work used computational resources of
supercomputer Fugaku provided by the RIKEN Center for Computational Science
(Project ID: hp210165) and Oakbridge-CX provided by the University of
Tokyo (Project IDs: hp200106, hp200130, hp210094) through the HPCI System
Research Project. Numerical computation was also carried out on PC
clusters in KEK Computing Research Center. This work was also supported
by computational time granted by the Greek Research and Technology Network
(GRNET) in the National HPC facility ARIS, under the project IDs
SUSYMM and SUSYMM2.




\bibliographystyle{JHEP}
\bibliography{cle-lorentz_ref}


\end{document}